\documentclass[11pt,fleqn]{article}
\usepackage{amssymb}

\usepackage{amsmath}

\usepackage{amsmath,amssymb,amsthm}

\begin{document}
\begin{center}
{\Large \textbf{Invariants for sets of vectors and rank 2 tensors,
and differential invariants for vector functions}}

\vskip 20pt {\large \textbf{Irina YEHORCHENKO}}

\vskip 20pt {Institute of Mathematics of NAS Ukraine, 3 Tereshchenkivs'ka Str., Kyiv-4, Ukraine} \\
E-mail: iyegorch@imath.kiev.ua
\end{center}

\vskip 50pt
\begin{abstract}
We outline an algorithm for construction of functional bases of absolute invariants
under the rotation group for sets of rank 2 tensors and vectors in the Euclidean space of arbitrary dimension. We will use our earlier results for symmetric tensors and add results for sets including antisymmetric tensors of rank 2.

That allowed, in particular, constructing of functional bases of differential invariants for vector functions, in particular, of first-order invariants of Poincar\'e algebra (invariance algebra of Maxwell equations for vector potential).
\end{abstract}

\section{The Problem}
It might seem that all problems related to description of
differential invariants were solved in the classical papers by
S.Lie or in multiple later papers. However, the author failed to find explicit 
results on functional bases of invariants of the rotation group for sets of arbitrary multidimensional rank 2 tensors and vectors in the existing literature. Although some of the presented results may seem obvious, their proofs are not always straightforward.

The only results widely available are fundamental bases of invariants for three dimensions or for one tensor and/or vector. It may be natural as specific applications, e.g. in mechanics, most often required only rotation invariants for three dimensions.

However, differential invariants became a popular subject recently, and many new aspects of this subject were studied. We will not give any comprehensive bibliography here, as even an approximately exhaustive reference list would exceed the length of
the paper itself. We listed only some papers as representatives of
certain aspects relevant to our specific research. 

Popularity of the subject and extensive applications of differential invariants e.g. in computer vision, image recognition and characterization of differential equations led the author to believe that systematic specific description of invariants for vectors and tensors in multidimensional space would be relevant. 

Here we will be using the standard language of the
symmetry analysis of differential equations and of description of invariants, see e.g. classical papers
by Lie \cite{RDI1:LieDI}, Tresse \cite{RDI1:Tresse};  books by Glenn \cite{Glenn}, Veblen \cite{Veblen}, Spencer \cite{RDI1:Spencer}, Ovsyannikov  \cite{RDI1:Ovs-eng} and Olver \cite{RDI1:Olver1}-\cite{RDI1:Olver3}.

This paper reviews and extends our previous research in 
\cite{RDI1:Yepreprintcomplexfields} and \cite{RDI1:FYeDifInvs}. In
\cite{RDI1:Yepreprintcomplexfields} we described bases of first order differential invariants of the Poincar\'e algebra for the vector potential (in four dimensions)
\begin{equation} \label{PA}
\partial_{x_\mu}, \ J_{\mu \nu}= x_\mu\partial_{x_\nu} - x_\nu\partial_{x_\mu} + J_{\mu \nu}= A_\mu\partial_{A_\nu} - A_\nu \partial_{A_\mu},
\end{equation}

\noindent
where $\partial_{x_\mu}$ designates the operator 
$\frac{\partial}{\partial {x_\mu}}$,
and in 
\cite{RDI1:FYeDifInvs} - second order differential invariants of the Euclid, Poincar\'e and Galilei algebras for sets of scalar functions. These results described differential invariants, but were actually based upon description of invariants for sets of vectors and symmetric tensors of rank 2. To the moment, the author have not found any earlier results on fundamental bases of invariants for sets of vectors and symmetric tensors of rank 2 for the rotation group. Well-known and obvious results are related to one tensor or a set of one vector and one rank 2 symmetric tensor.

As an arbitrary rank 2 tensor can be represented as a sum of a symmetric and an antisymmetric tensor (also of rank 2), description of invariants for such arbitrary tensors would require, in addition to already available functional bases for sets of symmetric tensors, also construction of invariants for antisymmetric tensors and for sets of symmetric and antisymmetric tensors.

In our general presentation we will consider vectors and tensors in the $n$-dimensional Euclidean space, with $n$ equal or larger than 4 - as the results for $n=3$ are more obvious, and the rank for the rotation algebra for $n=3$ may require specific consideration. The examples may include the Minkovsky space for three spatial and one time dimensions.

The new results in this paper are construction of functional bases of invariants for antisymmetric tensors and for sets of vectors, symmetric and antisymmetric tensors in the Euclidean space of arbitrary dimension.

In this paper we imply by invariants only absolute invariants.  

\subsection{Two ways to describe invariants}
Invariants of transformations may be described by functional bases of differential invariants of a particular order and by fundamental invariants.

\begin{enumerate} \itemsep=-1pt
\item Functional basis of invariants is a set of functionally independent invariants such that every invariant can
be represented as a function of invariants from this set.
\item Fundamental invariants are a set of invariants such that every invariant can be represented through differentiation of some functions of invariants in this set.
\end{enumerate}

Here we will construct functional bases of invariants.

\subsection{Variables and Transformations}
We will consider sets of vectors and tensors of the type
\[
U = (u_i), V = (v_{ik}), \quad i,k = 1, ..., n;
\]
and rotation operators in the $n$ dimensional Euclid space, $n \geq 4$

\[
J_{ik}= u^r_i\partial_{u^r_k} - u^r_k\partial_{u^r_i}+ v^s_{il}\partial_{v^s_{lk}} - v^s_{kl}\partial_{v^s_{li}}
\]

\section{Motivation}
\subsection{Differential Invariants of Vector and Tensor Functions}
Invariants of tensors we consider here can be used for description of differential
invariants of scalar, vector and tensor functions - the relevant sets of derivatives of vector and tensor functions may be
treated as vectors and tensors due to the forms of prolongations of the relevant rotation operators.

\begin{enumerate} \itemsep=-1pt
\item If we have scalar functions, first order differential invariants may be treated as invariants for sets of vectors.

\item If we have scalar functions, second order differential invariants invariants may be treated as  invariants for sets of symmetric tensors of rank two and vectors

\item If we have vector functions, first order invariants may be treated as   invariants for sets of arbitrary tensors of rank two and vectors

\item If we have vector functions, second order invariants may be treated as   invariants for sets of tensors of rank three that are symmetric on two indices, tensors of rank two and vectors.
\end{enumerate}

\section{General background}
The general background for the invariant theory may be found e.g. in \cite{RDI1:Olver3}.

\subsection{Definitions of Invariants}
{\bf Definition} A function
\[ 
F = F(U^r,V^s),
\]

\noindent where
$$
U^r = (u^r_i), V^s = (v^s_{ik}), \quad   i,k = 1, ..., n;
$$
are respectively sets of vectors and tensors of rank two,
is called
an absolute invariant (or simply invariant)
for the Lie algebra $L$ with basis
elements $J_{ik}$ if
$$
J_{ik}F=0.
$$

As in \cite{RDI1:FYeDifInvs}, it is easy to show that in principle it is sufficient to give all results for sets of two vectors and two tensors, as those may be easily extended for arbitrary numbers of vectors and tensors.

\subsection{Determining Equations}
The equations
$$
J_{ik}F=0,
$$
where $J_{ik}$ are infinitesimal operators of the Euclid rotations, are called determining equations for the invariants $F = F(U^r,V^s)$.

We see that the problem of finding invariants is actually a problem of finding solutions for a system of linear partial differential equations of first order.
Unfortunately, there is no theory helpful in the case of arbitrary number of equations, and we use some ad-hoc methods.

\subsection{Number of Invariants in the Functional Basis}
In our consideration, $V^s = (v^s_{ik})$, $i,k = 1, ..., n$, $s=1,2$ is a set of two arbitrary tensors of rank 2. We replace each tensor $V^s$ in our consideration by a pair of a symmetric and an antisymmetric tensors:

$$w^s_{ik}=v^s_{ik}+v^s_{ki},$$
$$y^s_{ik}=v^s_{ik}-v^s_{ki}.$$

The number of invariants in a fundamental set is the difference between the number
of independent variables and the general rank of the system of determining equations.

As an example, let us consider two vectors and two symmetric tensors of rank two.
In this case we have(see \cite{RDI1:FYeDifInvs})
$$ 2n+2\frac{n(n+1)}{2}=n^2+3n $$
independent variables.

The general rank of the system of determining equations in the case of rotational transformations is equal to the number of operators 
$$\frac{n(n-1)}{2}$$.

So we shall have 
$$2n+2\frac{n(n+1)}{2} - \frac{n(n-1)}{2}= \frac{7n +n^2}{2}$$ 
functionally independent invariants.

The case of one antisymmetric tensor in the multidimensional space appears to be quite cumbersome as determination of the rank of the rotation algebra for such tensor is not straightforward. For such tensor, we will have
$\frac{n(n-1)}{2}$ independent components, and $\frac{n(n-1)}{2}$ operators in the rotation algebra; however, the rank will be smaller as at least one invariant of the form $y_{ik}y_{ik}$ is obvious. We can prove that the rank for such tensor will be equal to

$$\frac{n(n-1)}{2}-n,$$

and we will have $n$ functionally independent invariants.

In the case of two vectors and two antisymmetric tensors of rank two (or just two arbitrary tensors of rank two), we will have

$$
2n+2n(n+1)
$$
variables, and whence
$$2n+2n(n+1) - \frac{n(n-1)}{2}= \frac{3n +n^2}{2}$$ invariants.

In the case of two vectors, two symmetric tensors and two antisymmetric tensors of rank two (or just two arbitrary tensors of rank two), we will have

$$
2n+2n(n+1)
$$
variables, and whence the number of functionally independent invariants is
$$2n+2n(n+1) - \frac{n(n-1)}{2}= \frac{9n +3n^2}{2}$$.

\subsection{Functional Bases of Invariants}
{\bf Theorem 1} \cite{RDI1:FYeDifInvs}. A functional basis of invariants of the rotational transformations for two vectors and two symmetric tensors in the $n$-dimensional space can be taken as follows:
\begin{gather}
w^1_{ii}, \ w^2_{ii}, \ u^r_i u^r_i, \ S_{a}(w^1_{ik}),\ S_{a}(w^2_{ik}), \ S_{ab}(w^1_{ik}, w^2_{ik}),\quad R_a(u^r_i, w^1_{ik}), \ r=1,2; \nonumber
\end{gather}

where
\begin{gather} \nonumber
S_{a}(w^r_{ik})=w^r_{i_1 i_2}w^r_{i_2 i_3}...w^r_{i_{a} i_{1}}
\end{gather}
We mean summation over the repeated indices from 1 to $n$.
In the lists of invariants $S_{a}$, $a$ takes the values from 1 to $n$.

\begin{gather} \nonumber
S_{ab}(w^1_{ik}, w^2_{ik})=w^1_{i_1 i_2}w^1_{i_2 i_3}...w^1_{i_{b} i_{b+1}}w^2_{i_{b+1} i_{b+2}}...w^2_{i_{a} i_{1}} \nonumber
\end{gather}

In the lists of invariants $S_{ab}$, $a$ takes the values from 2 to $n$ and $b$
takes the values from 1 to $a$.

\begin{gather} \nonumber
R_{a}(u^r_{i}, w^1_{ik})=u^1_{i_1} u^1_{i_a} w^1_{i_1 i_2}...w^1_{i_{a-1} i_{a}}
\end{gather}
In the lists of invariants $R_{a}$, $a$ takes the values from 1 to $n$.

{\bf Theorem 2}. A functional basis of the rotational transformations for two vectors and two antisymmetric tensors in the $n$-dimensional space can be taken as follows:
\begin{gather} \nonumber
S_{ab}(y^1_{ij}y^1_{jk}, y^2_{ij}y^2_{jk}),\quad R_a(u^r_i, y^1_{ij}y^1_{jk}); \\
i,k=1,...,n; \ a,b=1,...,n; \ r=1,2,\nonumber
\end{gather}

{\bf Theorem 3}. A functional basis of the rotational transformations for two vectors, two symmetric and two antisymmetric tensors in the $n$-dimensional space can be taken as follows:
\begin{gather} \nonumber
S_{ab}(y^1_{ij}y^1_{jk}, y^2_{ij}y^2_{jk}),\quad R_a(u^r_i, y^1_{ij}y^1_{jk}); \\ \nonumber
i,k=1,...,n; \ a,b=1,...,n; \ r=1,2,\nonumber
\end{gather}

\subsection{Idea of the Proof that Invariants in the Functional Basis
are Functionally Independent}
This part of the problem is the most difficult. Such proof can be done with
mathematical induction and different for different types of tensors.

Actually it is necessary to prove that a Jacobian of the basis of invariants is not zero,
that is a determinant of an $(N \times N)$-matrix, $N$ being the number of invariants in the functional basis.

If we assume that the set of invariants for the dimension $n$ is functionally independent,
we have to prove the same for the set of invariants for the dimension $n+1$.

Alternatively, such proof can be easily obtained from the proof in 
\cite{RDI1:FYeDifInvs}, if we consider $u_{k}y_{ki}$ as new vectors, and convolutions of $y_{ik}y_{kl}$ as new symmetric tensors.

To have the general results for tensors of rank two, we need to consider antisymmetric tensors and combination of symmetric and antisymmetric tensors.

\section{Example: Differential Invariants for the Vector Potential}
\subsection{Vector Potential Functions}
Here we present, following \cite{RDI1:Yepreprintcomplexfields}, an example of a functional basis of the first-order differential invariants for the vector potential functions

$$
A_\mu, \mu = 0, 1, ..., n,
$$

with respect to Poincar\'e algebra with the basis operators (\ref{PA}).
To look for the first-order differential invariants, we may consider invariants for the vector $(A_\mu)$ and two tensors:
a symmetric tensor $B_{\mu \nu}=\partial_{A_\nu} A_\mu + \partial_{A_\mu} A_\nu$, and an antisymmetric tensor
$L_{\mu \nu}=\partial_{A_\nu} A_\mu - \partial_{A_\mu} A_\nu$.

A functional basis of the first-order differential invariants may be taken as follows:
\begin{gather} \nonumber
A_\mu A_\mu, A_\mu B_{\mu \nu} A_\nu, A_\mu B_{\mu \nu}B_{\nu \alpha}A_\alpha,\\ \nonumber
A_\mu B_{\mu \nu} L_{\nu \alpha}A_\alpha, A_\mu L_{\mu \nu} L_{\nu \alpha}A_\alpha, \\ \nonumber
B_{\mu \mu}, B_{\mu \nu}B_{\mu\nu}, B_{\mu \nu}B_{\nu
\alpha}B_{\alpha \mu}, B_{\mu \nu}B_{\nu \alpha} B_{\alpha \beta}
B_{\beta \mu},\\ \nonumber
L_{\mu \nu}L_{\mu\nu}, L_{\mu \nu}L_{\nu \alpha} L_{\alpha \beta}
L_{\beta \mu}, L_{\mu \nu}L_{\nu \alpha}B_{\alpha \mu}, L_{\mu
\nu}B_{\nu \alpha} B_{\alpha \beta} L_{\beta \mu}, L_{\mu
\nu}B_{\nu \alpha} L_{\alpha \beta} B_{\beta \mu}. \nonumber
\end{gather}

In the case when  $\mu, \nu = 0,1,2,3$, the invariants shall depend on 20 components; the general rank of
the operators ${J_{\mu \nu}}$ is equal to 6, so we shall have 14
functionally independent invariants.

\section{Conclusion}
\subsection{Main ideas}

\begin{enumerate} \itemsep=-1pt

\item Description of invariants is needed for solution of other problems.

\item The task is split into solution of basic tasks (e.g. it is sufficient to consider two vectors/tensors
of each  type.

\item Description of invariants for special types of tensors is equivalent to finding
of conditional invariants \cite{cond-diff-inv} (we add other conditions to determining equations).

\end{enumerate}

\subsection{Further research}

Immediate follow-up
\begin{enumerate} \vspace{-1mm}\itemsep=-1pt

\item Description of relative invariants and covariant vectors/tensors

\item Description of differential invariants for various extensions
of the Euclid algebra

\item Description of invariants for various special tensors, e.g. will smaller number of independent components.
    
\item Finding of relations among functionally dependent invariants.

\item Description of invariants for nonlinear representations of the rotation group, see e.g. \cite{RDI1:IAY nonlin}
    
\end{enumerate}

These sets of invariants can be useful tools in other problems in studying
of partial differential equations.

\end{document}